# Efficient Spectrum Sharing in the Presence of Multiple Narrowband Interference

Demosthenes Vouyioukas

Department of Information and Communication Systems Engineering
University of the Aegean
Karlovassi 83200, Samos, Greece

e-mail: dvouyiou@aegean.gr

**Abstract.** *In this paper, we study the spectrum usage efficiency by applying wideband methods and systems to the existing analog systems and applications. The essential motivation of this work is to define the prospective coexistence between analog FM and digital Spread Spectrum systems in an efficient way sharing the same frequency band. The potential overlaid Spread Spectrum (SS) system can spectrally coincide within the existing narrowband Frequency Modulated (FM) broadcasting system upon several limitations, originating a key motivation for the use of the FM radio frequency band in many applications, encompassing wireless personal and sensors networks. The performance of the SS system due to the overlaying analog FM system, consisting of multiple narrowband FM stations, is investigated in order to derive the relevant bit error probability and maximum achievable data rates. The SS system uses direct sequence (DS) spreading, through maximal length pseudorandom sequences with long spreading codes. The SS signal is evaluated throughout theoretical and simulation-based performance analysis, for various types of spreading scenarios, for different carrier frequency offset ( f) and signal-to-interference ratios, in order to derive valuable results for future developing and planning of an overlay scenario.*

## 1. Introduction

Nowadays and future trends of wireless systems are mainly towards developing new wireless technologies so as to support higher data rates with minimum costs. This is achievable – apart from other – by demanding additional access to the spectrum. The spectrum scarcity is the most serious challenge facing the wireless industry today and it is only going to get worse during the next years. Major accomplishments have already been developed in the area of Cognitive Radio (CR) for spectrum sensing, spectrum sharing, dynamic spectrum access to primary and secondary users and between Radio Access Technology (RAT) [1],[2]. But, spectrum regulation still remains where many spectrum bands are unused and some of them are heavily congested allocated to specific applications and operators.

Enabling the provision of services and applications in an efficient way making use of the natural resource radio spectrum can be realized by introducing principles of wideband SS techniques in real existing regulation. The spectrum overlay between Direct Sequence Spread Spectrum (DSSS) systems and FM radio broadcasting services can increase spectral efficiency and communication capacity within a geographical area. Anti-jamming, anti-interference, privacy and low power spectral density are some of the advantages that the SS technique encompasses and strengthen their use to coexisting schemes. Both field tests and analyses have provided a perspective as to what the capabilities of such system are.

                                                                                          61

The mutual interference, which in specific conditions of high density becomes unavoidable, the estimation of interference noise serves as a measure for accepting their coexistence. This overlay concept has been demonstrated in both PCS band [3] and cellular band [4]. In the proposed work of [5], it has been proposed a CDMA network that is permitted to be overlaid on top of existing microwave narrow-band users that occupy a part of the spread-spectrum system bandwidth (BW). In fact, this is another advantage of CDMA over either TDMA or FDMA, because this will increase the overall spectrum capability. However, such an application must be considered carefully, because the CDMA (wide-band) users and the narrow-band users can interfere with each other. Analysis of Ultra-Wide-Band (UWB) systems reveals that signals are suitable for underlay communications and the design of these systems are very promising, where it is necessary the understanding of the effects of interference to and from narrow-band systems [6].

Simulcasting methods and techniques [7],[8] have been developed for simultaneous transmission of digital data with analog FM, over frequency bands in the 87.5 – 108 MHz range. Some of these new methods are compatible with the standards of analog FM for terrestrial broadcast, while others are designed for new frequency bands. The method of hybrid in band on channel (HIBOC) has been used, in order to transmit digital audio simultaneously with analog FM, using frequency-orthogonal analog and digital transmission over a total system bandwidth of 400 kHz.

Recent advances in wireless communications and electronics have enabled the development of various types of short-range wireless communication systems, such as WLANs, Bluetooth, High data rate WPANs, Low power WPANs, Body area networks (BANs) and RFID-technologies with a variety of applications (e.g., commercial, home, health, military, etc.). Most of these systems as well as the wireless sensors networks, utilize the spread spectrum technique for the physical layer [9]. The unusual application requirements of sensor networks make the choice of transmission media more challenging. One option for radio links is the use of industrial, scientific and medical (ISM) bands, which offer license free communication in most countries. These frequency bands suffer from limitations and harmful interference from existing applications and applications that may emerge in the future [10]. Another option for radio links is the use of the already occupied frequencies, in an overlay scenario utilizing the ultra low power transmission of the sensor nodes.

Exploitation of this spectrum overlay concept in the physical layer in the field of wireless personal and sensor networks, can increase communications capacity and spectral efficiency, but may cause the following types of interference: i) interference from the narrowband FM stations to the SS system and ii) interference from the overlaid wideband SS system on the FM receivers. The first is the scope of this paper. The later case was investigated and the correspondent results were presented in [11].

The performance of the Direct Sequence Spread Spectrum system is examined once overlaid on top of the existing radio band of multiple FM stations. The probability of error for various FM interfering conditions is derived for the case of single and multiple spread spectrum users. The simulation of the proposed configuration for the estimation of the error rate, aims for the best approach of a realistic system, as well as for the knowledge of the produced signals at the output of each module. Consequently, based on the analysis and according to mathematic analytical models of each module, the simulation of the whole system is realized, taking into consideration all the stimulated occurrences that are caused due to non-ideally operation of the passive and active elements. Moreover, the Additive White Gaussian Noise (AWGN) is taken into account for the Spread Spectrum system and also for each FM station.

In Section 2, the Spread Spectrum system is briefly described in the form used in this paper. The total interference analysis for single and multiple SS due to multiple narrowband interference users is performed in Section 3, while the FM power spectrum approximation is presented in Section 4 along with some of their properties in the context of closed-form mathematical expressions. Section 5 contains some indicative numerical results for single and multiple SS users, while in Section 6 simulation results for a single SS user are presented. Finally, a brief discussion and concluding remarks are given in Section 7.

## 2. Description of Spread Spectrum System

The spread spectrum signal is given by [12] as

$$s(t) = \sum_{k=1}^{K} s_k(t-\tau_k) = \sum_{k=1}^{K} \sqrt{2P_k} \, b_k(t-\tau_k) c_k(t-\tau_k) \cos\left(\omega_o t + \theta_k\right) \tag{1}$$

where $P_k$ is the received power of the $k$-th spread spectrum, $\tau_k$ is the time delay of the signal uniformly distributed into [0,T], $\theta_k$ is the phase angle uniformly distributed on [0,2$\pi$], $b_k(t)$ is the modulating digital signal of the $k$-th user given by $b_k(t) = \sum_{n=-\infty}^{\infty} a_n^{(k)} p_{T_b}(t-nT_b)$ where $\{a_n = \pm 1, -\infty < n < +\infty\}$ and $p_{T_b}$ is the rectangular pulse of $T_b$ duration, with P[$b_k(t)$=1]=P[$b_k(t)$=-1]=0.5, and $\omega_o$ is the carrier frequency of the signal. $c(t)$ is the spreading code given by $c_k(t) = \sum_{n=-\infty}^{\infty} c_n^{(k)} p_{T_c}(t-nT_c)$ where $c_k(t) \in \{\pm 1\}$ is one chip of random binary sequence $\{c_n\}$, which consists of independent identically distributed (i.i.d) random variables with equal probability, $T_c$ is the spreading code chip duration and $p_{T_c}$ is the chip waveform, assumed rectangular.

## 3. Interference Analysis

### 3.1   Single SS user

We consider the receiver model as depicted in Figure 1. The total received signal is corrupted by noise and interference

$$r(t) = s(t) + i(t) + n(t) \tag{2}$$

where $i(t)$ denotes the interference and $n(t)$ is the zero-mean white Gaussian noise. Assuming that code synchronization has been established and we have perfect channel estimation, the input to the demodulator is

$$r_1(t) = s(t)c(t) + i(t)c(t) + n(t)c(t) \tag{3}$$

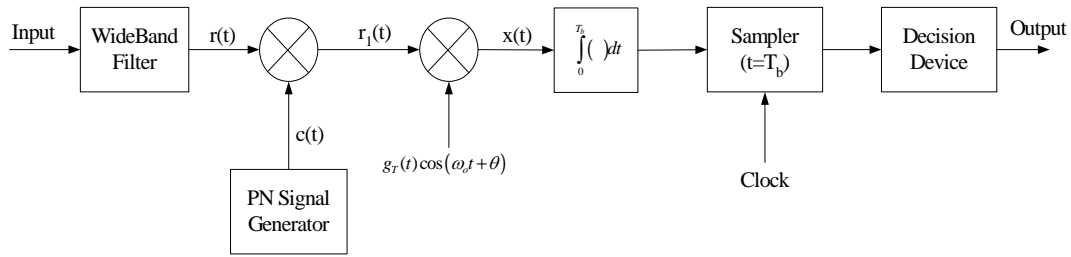

Figure 1: DS receiver model with BPSK modulation

The input quantity to the decision device, during the period $T_b$ of the data signal, is

$$Z_o = \int_0^{T_b} r_1(t)\cos\left(\omega_o t + \theta\right)dt = \int_0^{T_b}\left[\sqrt{2P}b(t)c(t)\cos\left(\omega_o t + \theta\right) + i(t) + n(t)\right]c(t)\cos\left(\omega_o t + \theta\right)dt =$$

$$= \sqrt{2P}\int_0^{T_b} b(t)c^2(t)\cos^2\left(\omega_o t + \theta\right) + \int_0^{T_b} i(t)c(t)\cos\left(\omega_o t + \theta\right)dt + \int_0^{T_b} n(t)c(t)\cos\left(\omega_o t + \theta\right)dt \quad (4)$$

We assumed that $f_o = \omega_o / 2\pi >> 1/T_b$, so that the double frequency term $\left(\cos^2\left(\omega_o t + \theta\right) \cong 1 + \cos\left(2\omega_o t\right) \cong 1\right)$ is negligible, suppressed by the bandpass filter following the demodulator. From equations (2), (3), (4) and the fact that $b(t)$ is constant over the period $T_b$ we have

$$Z_o \cong \sqrt{2P}\int_0^{T_b} b(t)\sum_{n=0}^{G_p-1} p_{T_c}^2\left(t - nT_c\right)dt + S_i + S_n = \pm\sqrt{2P}T_b + S_i + S_n,\ b(t) \in \left\{\pm 1\right\} \quad (5)$$

where

$$\int_0^{T_b}\sum_{n=0}^{G_p-1} p_{T_c}^2\left(t - nT_c\right)dt = G_pT_c = T_b \quad (6)$$

$$S_i = \int_0^{T_b} i(t)c(t)\cos\left(\omega_o t + \theta\right)dt \quad (7)$$

$$S_n = \int_0^{T_b} n(t)c(t)\cos\left(\omega_o t + \theta\right)dt \quad (8)$$

The code chip sequence $\left\{c_n\right\}$ can be modeled as a random binary sequence, comprises by statistically independent symbols with equal probability, is uncorrelated (white) and therefore $E\left[c_n c_m\right] = E\left[c_n\right]E\left[c_m\right]$ for n≠m and these conditions imply that $E\left[c_n\right] = 0$ and $E\left[c_n^2\right] = 1$. Perfect code, phase and symbol synchronization are assumed.

Substituting the spreading code into (7), we obtain

$$S_i = \sum_{n=0}^{G_p-1} c_n J_n \quad (9)$$

where

$$J_n = \int_{nT_c}^{(n+1)T_c} i(t)p_{T_c}\left(t - nT_c\right)\cos\left(\omega_o t + \theta\right)dt \quad (10)$$

Likewise

$$S_n = \sum_{n=0}^{G_p-1} c_n N_n \quad (11)$$

where

$$N_n = \int_{nT_c}^{(n+1)T_c} n(t)p_{T_c}\left(t - nT_c\right)\cos\left(\omega_o t + \theta\right)dt \quad (12)$$

The equation (5) can also be written as $S \cong S_o + y_i$ where $S_o$ is the desired signal and $y_i = S_i + S_n$ denotes the total additive interference plus Gaussian noise. The first term of the right-hand side of the equation is deterministic and its value is given by equation (5). The terms of $S_i$ are independent random variables with zero mean value uniformly bounded and $\text{var}\{S_i\} \rightarrow \infty$ with $G_p \rightarrow \infty$, so applying the Central Limit Theorem (CLT) [14] implies that $S_i$ converges in a Gaussian distribution with zero mean value and variance 1. Consequently, the distribution of $S_i$ is almost Gaussian when $G_p$ is large. We must underline, that for the present analysis long sequences have been utilized with large processing gain. Therefore, the second term is a summation of two zero-mean independent Gaussian variables, indicating that $y_i$ will have an almost zero-mean Gaussian distribution and variance given by $\text{var}\{y_i\} = \text{var}\{S_i\} + \text{var}\{S_n\}$. Based on CLT and Gaussian approximation, the probability of error depends on the statistical characteristics of the interference. The final probability of error will be given by [14]

$$P_e = \frac{1}{2}\text{erfc}\left[\sqrt{\frac{E_b T_b}{\text{var}\{S_i\} + \text{var}\{S_n\}}}\right] \qquad (13)$$

and the total signal-to-noise ratio at the output of the correlator by

$$SNR_D = \frac{E^2[S]}{\text{var}\{S_i\} + \text{var}\{S_n\}} \qquad (14)$$

where $E_b = PT_b$ is the received bit energy $b(t)$, $P$ is its mean power and *erfc* is the complementary error function. Based on the previous definitions, the variance of the variable $y_i$ is

$$\sigma_{y_i}^2 = \text{var}\{y_i\} = \text{var}\{S_i\} + \text{var}\{S_n\} = E[S_i^2] + E[S_n^2]$$
$$= G_p E[J_n^2] + G_p E[N_n^2] \qquad (15)$$

Suppose that   is an independent random variable uniformly distributed in over $[0, 2\pi]$. The stationary of $n(t)$ and a change of variables implies that the variance of $N_n$ is

$$E[N_n^2] = \frac{N_o T_c}{4} \qquad (16)$$

where $\dfrac{N_o}{2}$ is the two-sided noise power spectral density (PSD) and we assumed that $f_o = \omega_o / 2\pi >> 1/T_c$.

For the calculation of the variance of interference, following the noise procedure results to [13]

$$E[J_n^2] = \frac{1}{2}T_c \int_{-T_c}^{T_c} R_i(\tau) R_p(\tau)\cos\omega_o\tau d\tau \qquad (17)$$

where $R_i(\tau)$ is the autocorrelation of $i(t)$, $R_p(\tau)$ is the autocorrelation function of the PN sequence.  he limits of the integral can be extended to infinity because the integrand is truncated. Because $R_i(\tau)$ is an even function, the convolution theorem and the known *Fourier* transform of $R_p(\tau)$, yield the following [15]

$$E[J_n^2] = \frac{1}{2}T_c^2 \int_{-\infty}^{\infty} S_i(f)\sin c^2[(f-f_o)T_c]df \qquad (18)$$

where $S_i(f)$ is the PSD of the interference after passage through the wideband bandpass filter. If $S_i^{'}(f)$ is the PSD of the interference at the input of the wideband bandpass filter and $H(f)$ is its transfer function, then $S_i(f) = S_i^{'}(f)|H(f)|^2$. Suppose that the effects of the wideband filter and the integration over negative frequencies are negligible ($f_o \gg 1/T_c$), then

$$E\left[J_n^2\right] = \frac{1}{2}T_c^2 \int_{f_i-W_i/2}^{f_i+W_i/2} S_i(f)\sin c^2\left[\left(f-f_o\right)T_c\right]df \qquad (19)$$

where we have assumed that the bandwidth of the interference is $W_i \leq W \cong 2/T_c$. The PSD of the interference $S_i(f)$ is given by equation (35) at the next section and the total probability of error is given by

$$P_e = \frac{1}{2}\text{erfc}\left[\frac{E_b}{N_o + 2\dfrac{T_b}{G_p}\displaystyle\int_{f_i-W_i/2}^{f_i+W_i/2} S_i(f)\sin c^2\left[\left(f-f_o\right)T_c\right]df}\right]^{\frac{1}{2}} \qquad (20)$$

### 3.2   Multiple SS users

The scope of this section is to examine the performance of the Direct Sequence Spread Spectrum system in the case of multiple spread spectrum active users and to access the bit error probability in the presence of analog FM interference and multiple access interference (MAI), using the simplified improved Gaussian approximation.

We assume communication through point-to-multipoint transmission, so multiple access interference is considered with perfect code, phase and symbol synchronization, particularly for downlink transmissions.

The total received signal in the input of the desired spread spectrum receiver is

$$r(t) = \sum_{k=0}^{K-1} s_k(t-\tau_k) + i(t) + n(t) = \sum_{k=0}^{K-1}\sqrt{2P_k}\,b_k(t-\tau_k)c_k(t-\tau_k)\cos(\omega_o t + \phi_k) + i(t) + n(t) \qquad (21)$$

where $s_k(t)$ is the $k$-th user's transmitted signal, $b_k$ is the data sequence for user $k$, $c_k$ is the spreading sequence for user $k$, $_k$ and $_k$ are random delay and carrier phase terms relative to the desired reference user $0$, $P_k$ is the received power of user $k$,      is the carrier frequency of the signal, $i(t)$ the interfering analogue FM signal and $n(t)$ denotes the zero mean white Gaussian noise of the channel.

The received signal contains the desired user, the $K$-$1$ undesired users and the interfering FM signal plus noise and all signals are mixed down to baseband, multiplied by the PN sequence of the desired user $0$ and integrated over a bit period $T_b$.

The decision statistic of the receiver for user $0$ is:

$$Z_o = \int_0^{T_b} r(t)c_o(t)\cos(\omega_o t + \theta)\,dt \qquad (22)$$

which may be expressed as

$$Z_0 = I_0 + I_{MAI} + I_{FM} + I_n \qquad (23)$$

where

$$I_o = \sqrt{2P_o}\int_0^{T_b} b(t)c_o^2(t)\cos^2(\omega_o t)dt \qquad (24)$$

$$I_{MAI} = \int_0^{T_b}\sum_{k=1}^{K-1}\sqrt{2P_k}\,b_k(t-\tau_k)c_k(t-\tau_k)\cos(\omega_o t + \phi_k)c_o(t)\cos(\omega_o t)dt \qquad (25)$$

$$I_{FM} = \int_0^{T_b} i(t)c(t)\cos(\omega_o t + \theta)dt \qquad (26)$$

$$I_n = \int_0^{T_b} n(t)c(t)\cos(\omega_o t + \theta)dt \tag{27}$$

where $I_0$ is the desired contribution to the decision statistic from the desired user $k=0$, $I_{MAI}$ is the multiple access interference, MAI, from other users, $I_{FM}$ is the contribution of the interfering overlaid FM signal and $I_n$ is the thermal noise contribution. The contribution of the FM interference and noise term is investigated and analyzed previously ($S_i$ and $S_n$ respectively).

Since the performance of the Spread Spectrum system under investigation is assumed to be interference limited, which means that the FM interfering signal is the dominant interference component, the number of simultaneous users is limited (relative small $K$). Hence it is essential for the calculation of MAI to use an approximation, which is accurate for relative small values of $K$.

For the calculation of the Multiple Access Interference contribution and finally the overall performance of the system, the simplified expression of the Improved Gaussian Approximation (SIGA) is used.

### 3.2.1 Performance of the overlaid system based on SIGA

The simplified improved Gaussian approximation to the bit error rate is based upon the premise that the MAI converges to a Gaussian random variable as the number of chips per data bit $N$ becomes large and for any $K$ [16]. Based on the expansion of differences method [17], the SIGA requires only that the mean $\mu$ and the variance $\sigma^2$ of $\psi = var(I_{MAI})$ to be determined.

Based on the Central Limit Theorem and the fact that all interference components of the decision statistic are assumed to be independent random variables with distribution, which converges to Gaussian for large N, the variance of the decision statistic, is

$$var(Z_0) = var(I_{MAI}) + var(I_{FM}) + var(I_n) \tag{28}$$

The contribution of noise and analog FM interference to the decision statistic is analyzed previously. For the case of perfect power control such that all users have identical power levels and these power levels are not random, the mean $\mu_\psi$ and the variance $\sigma_\psi^2$ of $\psi = var(I_{MAI})$, are given by [18]

$$\mu_\psi = \frac{G_p T_c^2}{6} P(K-1) \tag{29}$$

$$\sigma_\psi^2 = (K-1)\frac{T_c^4}{4}P^2\left[\frac{23 G_p^2}{360} + G_p\left(\frac{1}{20} + \frac{K-2}{36}\right) - \frac{1}{20} - \frac{K-2}{36}\right] \tag{30}$$

and the probability of error is given by [18]

$$P_e = E\left[Q\left(\sqrt{\frac{E_b T_b}{2\left[y + \frac{N_0 T_b}{4} + \frac{1}{2}T_b T_c S_i(f)\right]}}\right)\right] \tag{31}$$

The simplified bit error probability expression for the case of MAI [16], analog FM and noise interference is given by

$$P_e^{(SIGA)} \approx \frac{1}{3} erfc\left(\sqrt{\frac{E_b T_b}{4(\mu_\psi + I_{eq})}}\right) + \frac{1}{12} erfc\left(\sqrt{\frac{E_b T_b}{4(\mu_\psi + \sqrt{3}\sigma_\psi + I_{eq})}}\right)$$

$$+ \frac{1}{12} erfc\left(\sqrt{\frac{E_b T_b}{4(\mu_\psi - \sqrt{3}\sigma_\psi + I_{eq})}}\right) \tag{32}$$

where

$$I_{eq} = var\{I_n\} + var\{I_{FM}\} = \frac{N_0 T_b}{4} + \frac{1}{2} T_b T_c S_i(f) \tag{33}$$

and $S_i(f)$ is the closed-form FM power spectrum approximation given by equation (35) at the next section, where the FM power spectrum aggregation is calculated over the bandwidth of all the under-utilized FM stations.

## 4. FM Power Spectrum Approximation

In order to obtain accurate results for the BER of equations (20), (32) and (33), the interference power of the FM spectrum at the input of the digital receiver is required in closed-form mathematical expression. The required power spectral densities of the interference FM signals are presented as a result of laboratory measurements and mathematical derivations.

The FM signal is generally a random process and consequently can be only statistically presented. Specifically, to characterize the power spectral density of the FM signal we must measure the values of statistical parameters, such us mean, maximum, minimum and median. The measurement configuration consists of a spectrum analyzer, a biconical antenna and a laptop computer. The program in the computer controls the spectrum analyzer through HP-IB port. The spectrum analyzer parameters such as resolution bandwidth, video bandwidth, frequency span and sweep time are assigned through the Laptop PC. Three kinds of stations were recorded depending on the transmitted information; one that transmits only music, one that transmits only speech (voice) and one that transmits the combination of these two.

The statistical evaluation of the previous measurements and the derivation of a closed-form mathematical expression have been studied in [13]. Further statistical processing showed that the Gaussian fit for the estimation of the PSD of the FM channel responds very well according to several conditions of the transmitted data. The equation is given by the following expression:

$$\hat{G}_m(f) = y_o + A \exp\left[-\frac{(f - f_m)^2}{2\omega^2}\right] \tag{34}$$

where the parameters $y_o$ is a constant near the noise threshold, $A$ is the amplitude constant, $f_m$ is the center frequency,    is the standard deviation and $R^2$ denotes the regression analysis of the fit data. All the Gaussian fit parameters are given in Table 1, while Figure 2 depicts the mean recorded values and the Gaussian curve fit.

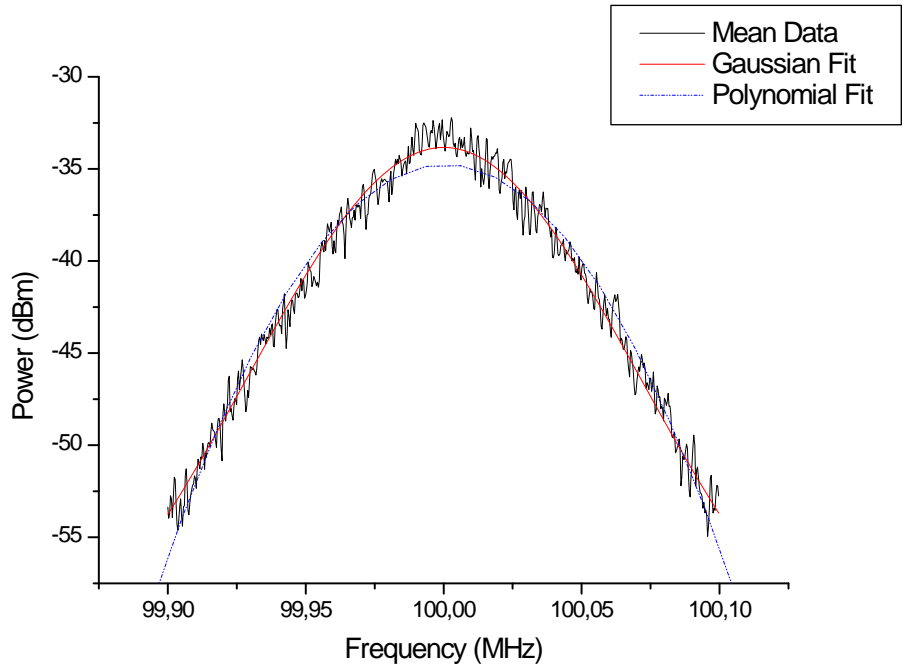

Figure 2: Mean value of a FM station and its spectrum approximation

Table 1: Gaussian fit parameters

| Parameters | Values |
| --- | --- |
| $y_o$ | -94.20817 dBm |
| $A$ | 58.92433 dBm |
| $f_m$ | 100 MHz |
|  | 0.09697 MHz |
| $R^2$ | 0.985 |

The total FM interference from all FM stations will be the sum of their PSD's

$$S_i^{'}(f) = \sum_{m=1}^{M} \hat{G}_{m(f)} = My_o + A\sum_{m=1}^{M} \exp\left[\frac{\left[f - f_1 - (m-1)\Delta f_m\right]^2}{2\omega^2}\right] \qquad (35)$$

where $M$ is the number of the FM stations, $y_o$, $A$ and  are given in Table 1, $f_1$ is the frequency of the first FM station and $f_m$ is the carrier separation between two subsequent stations.

## 5. Numerical Results

For the calculation of the probability of error for single SS user interfered by multiple FM stations, we applied the CLT for the estimation of the total power spectrum of all FM stations, implementing the mean value by $f_m = \dfrac{f_1 + f_2 + ... + f_M}{M}$ and standard deviation $\omega_m = M\omega$. We assumed that all the FM stations, $M$, have the same standard deviation and thus power. For best results and worst case scenario, we must consider as many stations as they can fit in the FM frequency band of 20 MHz.

Suggestively, we present the results in Figure 3 to Figure 5 applying equation (20) for a typical value of processing gain, e.g. $G_p = 50 = 17\,\text{dB}$, and for a typical carrier separation between the FM stations, e.g. $f_m$=500kHz. The results show the probability of error versus $E_b / N_o$ for various signal-to-interference ratios P/P$_J$, chip rate $f_c$, FM stations $M$ and frequency differences $f$ between the two systems. The results are also compared using the probability of error for the ideally BPSK performance $\left( P_e = \dfrac{1}{2}\,erfc\left(\sqrt{\dfrac{E_b}{N_o}}\right) \right)$.

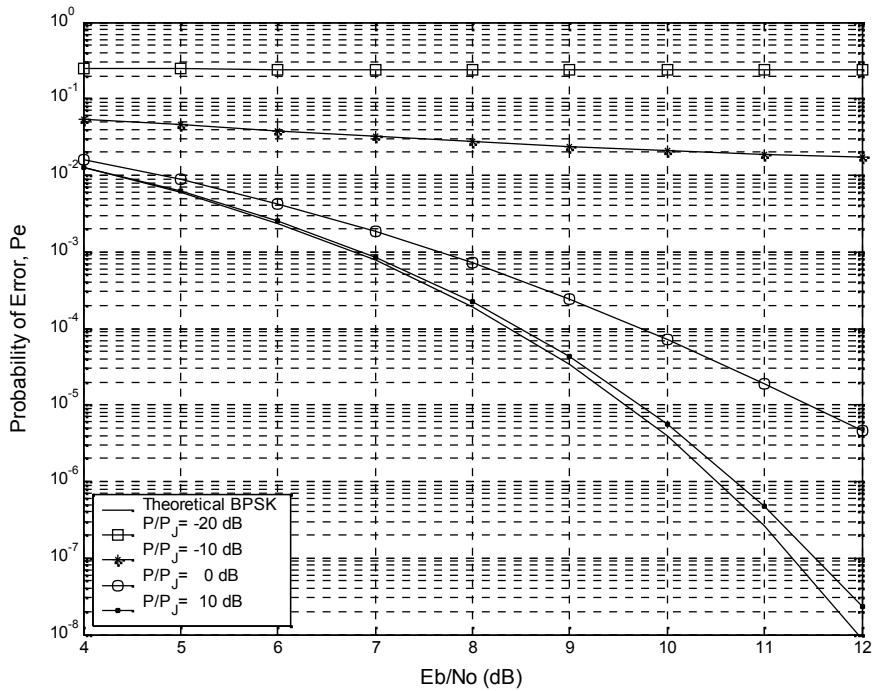

Figure 3: Bit error probability for $f$=0, $f_c$=10Mbps, M=40 and $W_i$=20MHz

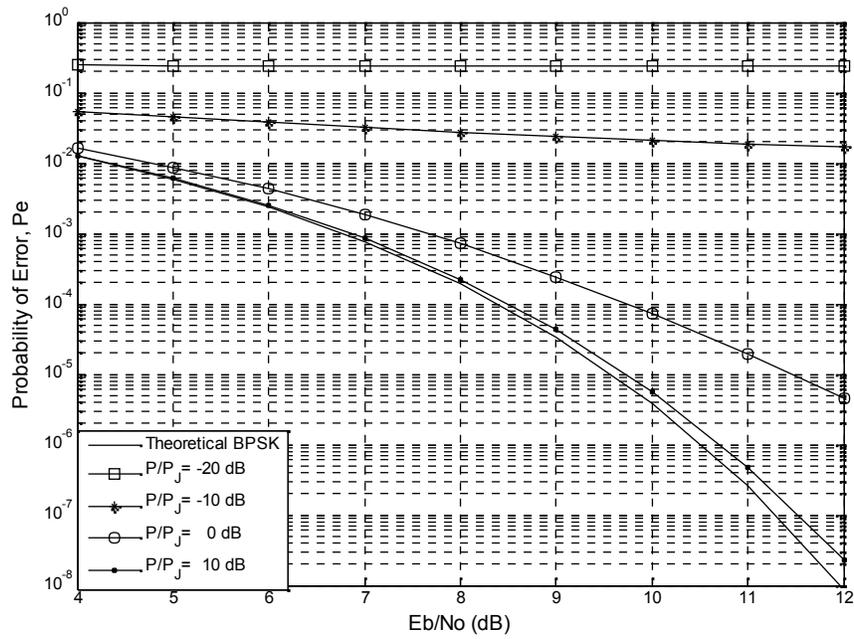

Figure 4: Bit error probability for $f = f_c/2$, $f_c = 1$Mbps, M=40 and $W_i = 20$MHz

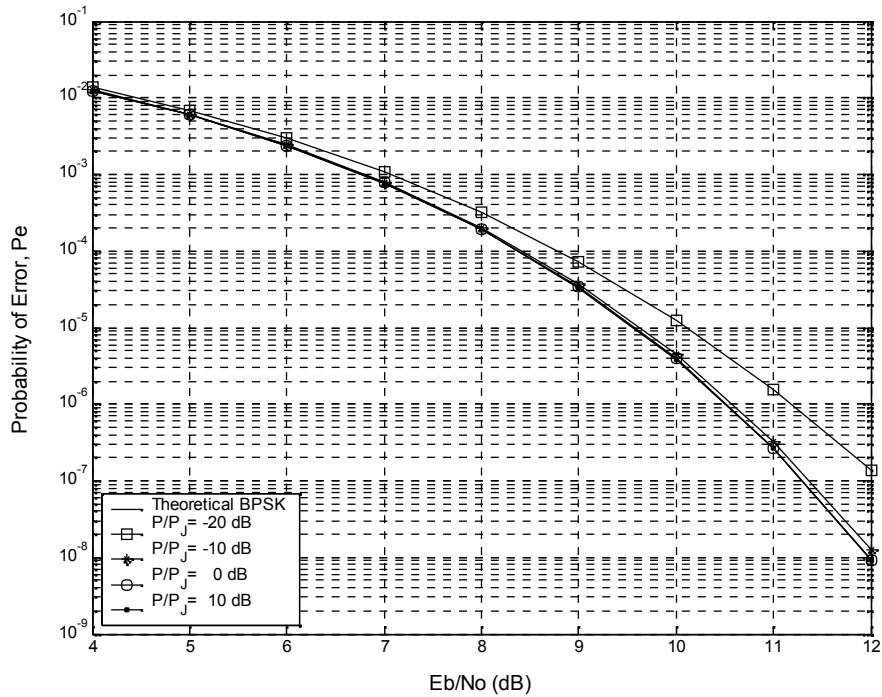

Figure 5: Bit error probability for $f = f_c$, $f_c = 5$Mbps, M=10 and $W_i = 5$     z

As can be shown from Figures 3-5, there is dependency between the carrier spacing of these two systems, the chip rate of the SS system and the number of FM stations in the frequency band. The worst case performance of the SS system is occurring when the two systems has identical center frequency ( $f$=0) and the number of FM stations are large. While the frequency offset $f$ between the center frequencies of these two systems is kept as large as possible and furthermore the chip rate of the SS system has maximum values, in that case the system performs ideally, even for great extent of interference.

The performance of the SS system utilizing multiple users is derived by using equations (32) and (33). In Figure 6, the bit error probability as a function of active Spread Spectrum users is presented, for $M$=40 interfering FM stations, $E_b / N_o$ =5dB and 10dB, processing gain $G_p = 50 = 17 \mathrm{dB}$ , chip rate $f_c$=10Mbps, frequency offset between the two systems $f$=0 and signal-to-interference ratios $P/P_J$= -20dB, -10dB and 0dB. Accordingly, Figure 7 depicts the bit error probability as a function of active Spread Spectrum users, for $M$=10 interfering FM stations, $E_b / N_o$ =5dB and 10dB, processing gain $G_p = 50 = 17 \mathrm{dB}$ , chip rate $f_c$=10Mbps, frequency offset between the two systems $f$=5MHz and signal-to-interference ratios $P/P_J$= -20dB, -10dB and 0dB. In Figure 8 the bit error probability as a function of active spread spectrum users is presented, for $M$=10 interfering FM stations, $E_b / N_o$ =5dB and 10dB, processing gain $G_p = 50 = 17 \mathrm{dB}$ , chip rate $f_c$=10Mbps, frequency offset between the two systems $f = f_c /2$ and signal-to-interference ratios $P/P_J$ =  -20dB, -10dB and 0dB.

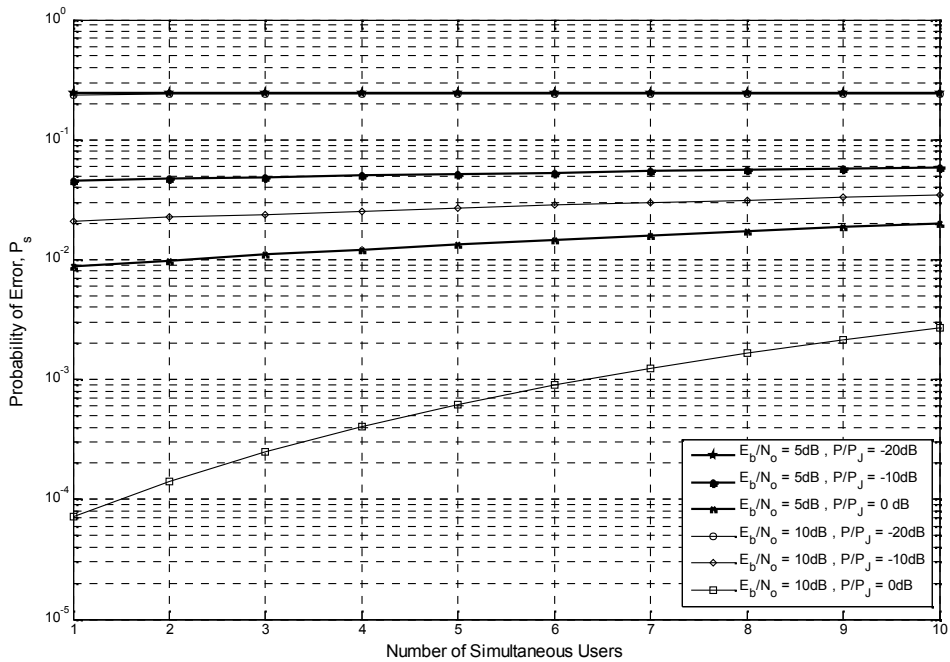

Figure 6: Bit error probability for $f$=0, $f_c$ =10Mbps, M=40 and $G_p$=50

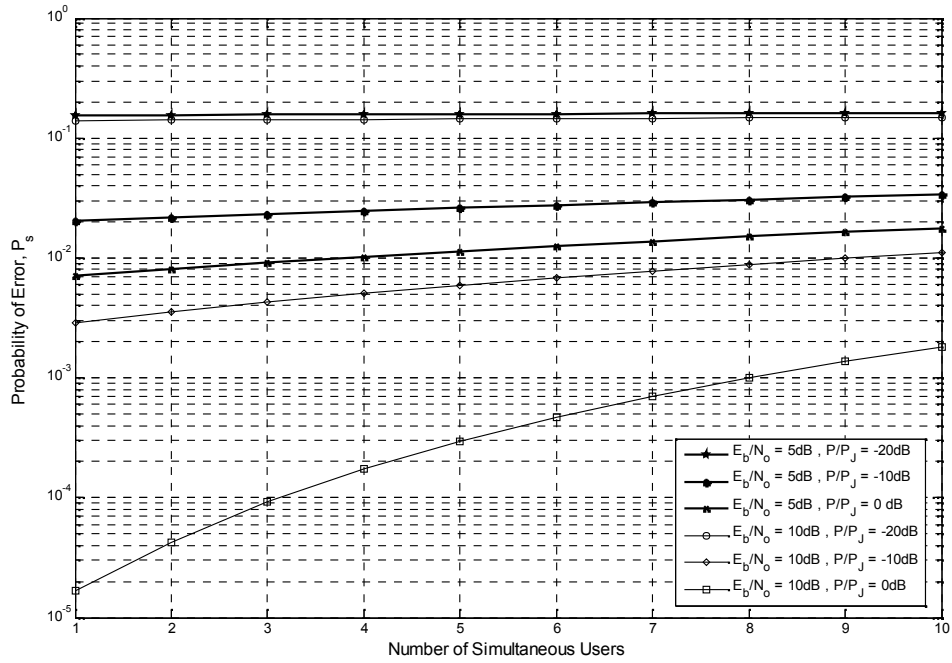

Figure 7: Bit error probability for $f$=5MHz, $f_c$=10Mbps, M=10 and $G_p$=50

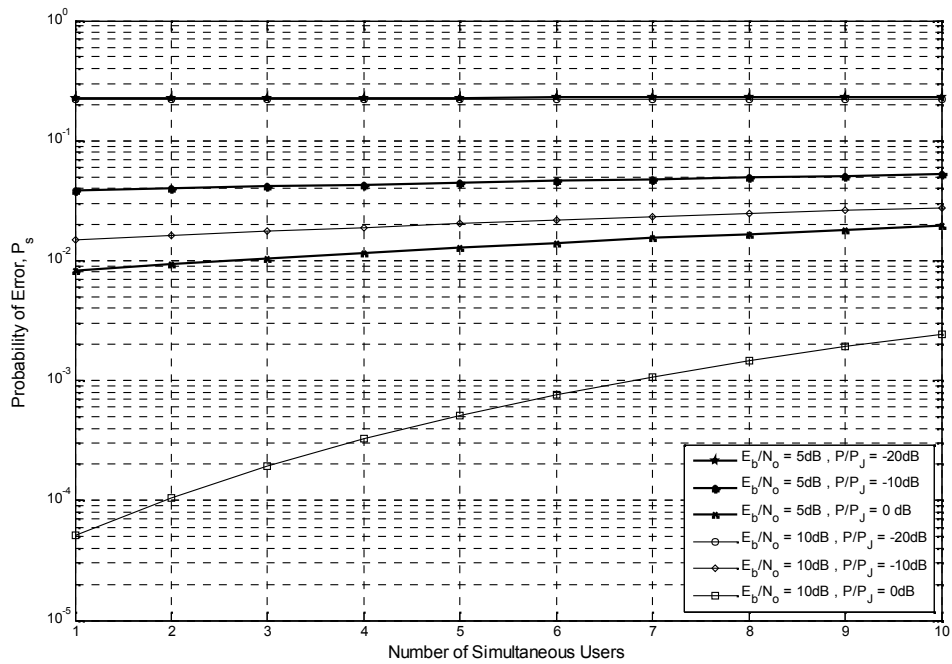

Figure 8: Bit error probability for $f$= $f_c$/2, $f_c$=10Mbps, M=10 and $G_p$=50

The results showed that there is a huge dependence between the spreading codes, the frequency distance between the two systems and the amount of interference, while the probability of error deteriorates when the number of simultaneous users is considerably increased.

# 6. Simulation Results

The reported objectives of this study will be supported with the help of an advanced design simulation system (Agilent-ADS) and evaluated in a realistic environment. The approach is based initially on theoretical level and then on simulation level. During the simulation, the effect of all factors that have analytical expression is verified, as well as these that do not have, trying to evaluate valid situations and systems [13].

The rationale of the simulation system consists of several steps. The first step comprises of the simulation of the spread spectrum transceiver, where the configuration for BPSK modulation is taken into consideration. Specifically, for the receiver's design, the correlator model of receiver has been noticed, as already discussed in Section 2 and presented graphically in Figure 2. A calibration of the system has been performed, in order to ensure for the accuracy of the process in terms of theoretical SNR vs. $E_b/N_o$.

The second step is the simulation of each FM station separately, so as to correspond, as much as possible, to actual conditions. Therefore, it has to be taken into account all the features that completely characterize the emitted frequency modulated signal (pre-emphasis, deviation, etc.). Likewise, the total FM spectrum is reproduced upon the recommendations of ITU for FM transmission [19], where the spectral output appears like the measured one of Figure 2. Subsequently, this module of FM station is characterized as independent subunit and the RF sum of all subunits provides the total multiple narrowband interference in the proposed system.

During the simulation process of signal transmission of SS system, we dealt with the total radio path, which is characterized by the multipath transmission, the propagation environment, the movement of receiver and the heights of the antennas. A typical propagation model for urban environment was selected, because the propagation channel effect wasn't the focal mission of this study. The proposed model for the radio path channel was incorporated into the process so as the final system of transmitter and receiver to be calibrated, without the FM interference, but with the addition of white noise, leading to a produced bit error rate similar to the theoretical ones.

Integrating all FM subunits constitutes the final step of the simulation configuration, which comprises the total model of FM stations in the final system of transmitter and receiver, illustrating the multiple narrowband interference to the system. Several number of interfered narrowband FM stations are considered during the simulation procedure.

The total performance of the SS system is measured through the estimation of the error rate at the output of the SS receiver. In order the bit error rate measurements to become more precise and efficient, the model of SS system should be regulated so as to include all the effects of their individual elements, apart from the receiver's noise. The control of system noise will become from a controlled noise source. This is what regulates also the quantity $E_b/N_o$.

The accuracy of BER measurements is actually only estimation and not the precise measurement, which their accuracy depends on the quantity of received samples. It is also reported as Monte Carlo approach. Generally, the BER estimation is usually very time-consuming process for practical applications. Typically, the technique of Importance Sampling [20],[21] is used, in order to substantially decrease the number of required samples. According to this technique during the simulation, the characteristics of the system are differentiated so as to vary the probability density function. Eventually, it is the preferable technique for the enhancement of BER measurements.

For the implementation of the simulation system certain constant parameters of the units and subunits have been used, in conjunction with variables ones. Because of the complexity and the large calculating time of the simulation, we varied only some of the parameters of the SS and FM system. At Table 2, the constant and the variable parameters of the simulation system are depicted. Prior to the FM interference employment, a calibration of the system has been performed, in order to ensure for the accurate feasibility of the process (theoretical SNR vs. $E_b/N_o$).

Table 2: Simulation parameters

| *Constant* | *Variable* |
| --- | --- |
| Processing Gain (G=17dB) | SS Power |
| SS Bandwidth 20MHz ($f_c$=10Mbps) | FM Power (each station) |
| Frequency separation between two consequently FM carriers (500kHz) | $E_b / N_o$ |
| Signal-to-Noise Ratio for SS system and each FM station (SNR = 10dB) | Variance of the error estimator |
| Propagation channel and antenna gain | Frequency separation ( f) between the SS system and the central frequency of FM band |

Figure 9 and Figure 10 depict the results of the simulation regarding the bit error rate at the output of the correlator for various ratios of wanted – to – unwanted signals (CIR) and frequency separation ( f) between the SS system and the middle frequency of the FM band. At this point, firstly we must mention that the calculation of the power of the interference includes the power from all the FM stations that are contained in the predetermined by the ITU frequency band and secondly, the variance of the error estimator had the value of 0.01. The specific value was satisfactory enough, but particularly aggravating for the simulation system. Because of the time-consuming simulation process, we tried to depict results comparable to the theoretical ones, like i.e. the numerical results of the analytical expressions in Section 5.

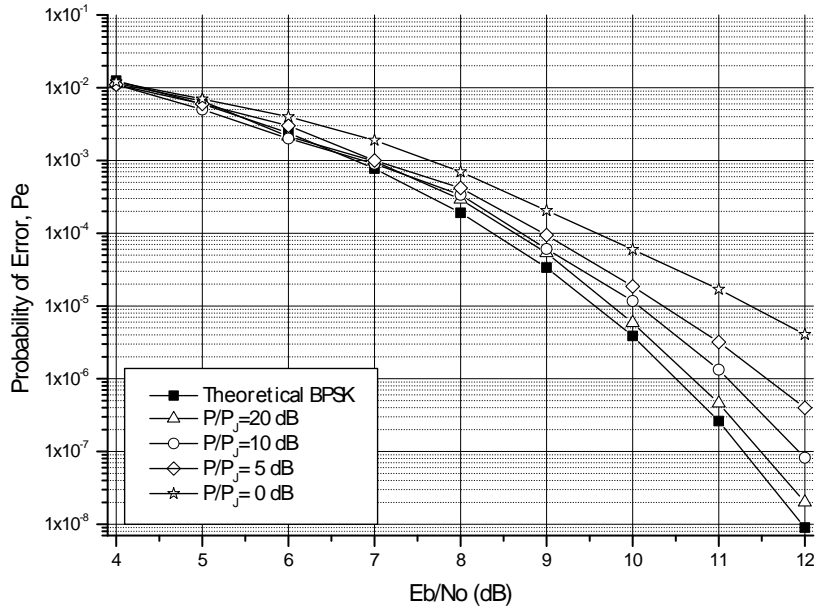

Figure 9: Bit error probability for $f$=0MHz, $f_c$=10Mbps, M=10 and $f_m$=500kHz

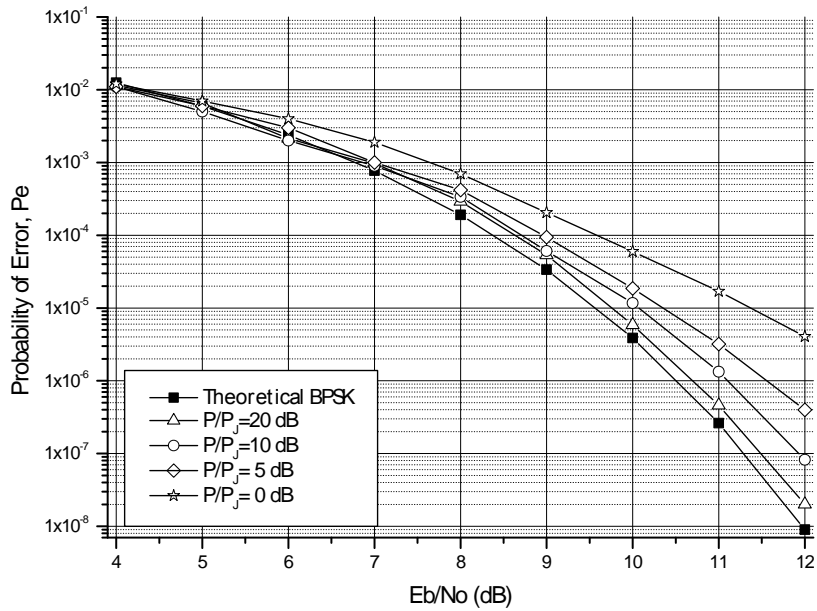

Figure 10: Bit error probability for $f$= $f_c$/2, $f_c$=10Mbps, M=10 and $f_m$=500kHz

# 7. Discussion and Conclusions

In this study, an efficient spectrum sharing approach has been proposed increasing the usage of real existing regulated spectrum by applying a wideband spread spectrum system. In particular, the investigation of spectrum overlay of a Direct Sequence Spread Spectrum system on the existing narrowband FM broadcasting system has been presented. We considered the implementation of a wideband SS system consisting of single and multiple users, situated on top of an existing conventional multiple narrowband FM system utilizing the 87.5 – 108 MHz band, causing interference on the SS system.

We examined the amount of multiple analog FM interference, which would not cause excessive degradation in the co-located SS system, through extensive interference analysis and simulation results. The conducting research was based on simulation of communication systems; it is critical to validate the system under development to assure that the results are indicative of the systems operation in a usable environment. The parameters of these two systems that had to be taken into consideration were several enough so as to designate the justification of this study. No channel coding has been used because our primary endeavour was to ensure cooperation between these two systems. The results of this study provide all the preliminary - but necessary - limitations attempting to spectrally coexist two dissimilar systems.

The evaluation of error probability based on simulation procedure provides interesting results that are, in some cases, close to the results obtained from the theoretical approximation in section 4, and when Figures 9 and 10 with Figures 3 and 4 are compared respectively. This implies that the assumptions and approximations proposed in section 3 and 4 and introduced in analytical expressions for the probability of error, are acceptable and barely affect the obtained results. Divergences between theoretical and simulation results are observed for high values of signal-to-interference ratio $P/P_J$, where the simulation procedure gives more strict interference levels.

Following the Central Limit Theorem, the Gaussian approximation for the derivation of the total FM interference and the simplified expression of the Improved Gaussian Approximation (SIGA) for multiple SS users, we conclude to the following results:

- The worst case for the probability of error occurs when the two systems have identically frequencies.
- The frequency difference $f$ has minimal impact to the performance of the system, while the number of FM stations is sufficiently large.
- As the total number of the FM stations is kept large considering small frequency separation $f_m$, the curves of bit-error-rate are improved.
- The system performance is highly degraded, especially when the number of active simultaneous users is increasingly rapidly.
- For more than roughly ten users the degradation of the performance reaches a threshold over which additional users have no impact on the system performance for the same FM interfering conditions.

In general, this study gives a key motivation for the exploitation of a different radio frequency band for the physical layer, than the ones the standards provide, in wireless personal and sensors networks. Extended research activities are currently undertaken by the authors in the area of a detailed SS system optimisation, in terms of coding and interference cancellation techniques, in order to establish the maximum allowable bit rate for the data signal. However, the results depend also on the choice of broadcasting parameters, such as the acceptable interference level in the demodulated analog FM audio signal (SNR degradation), as presented in [11].